Artículo de Investigación

# Mapa científico de la Inteligencia Artificial en Comunicación (2004-2024)

## Scientific Map of Artificial Intelligence in Communication (2004-2024)

**Carmen Gálvez**: Universidad de Granada, España.
cgalvez@ugr.es

**Cómo citar el artículo:**

**Resumen:**
**Introducción**: La Inteligencia Artificial (IA) está teniendo un impacto en el campo de la comunicación, provocando cambios transcendentales en la elaboración y en el consumo de la información. El objetivo de este trabajo fue analizar las áreas temáticas de la IA más influyentes en el campo de la comunicación a partir de la literatura científica. **Metodología**: Se seleccionaron 996 referencias indexadas en *Web of Science* entre 2004-2024, se realizó un análisis bibliométrico de co-palabras y se aplicaron técnicas de visualización para construir mapas científicos. **Resultados:** Las áreas temáticas más relevantes fueron la datificación, la vinculación de la IA con los medios sociales y el periodismo digital. Se identificó el área emergente de la IA generativa, vinculada a los nuevos modelos de IA, como ChatGPT, diseñados para generar contenido en forma de texto escrito, audio, imágenes o vídeos. Otra área temática emergente fue la repercusión de China en el uso de IA en comunicación. **Discusión**: A pesar del impacto de la IA en el ámbito de la comunicación, el campo todavía se encuentra en vías de estructuración, con pocos temas consolidados. **Conclusiones**: Este estudio permitió identificar las áreas temáticas del campo estudiado, así como la detección de las tendencias emergentes.

**Palabras clave:** inteligencia artificial; comunicación; bibliometría; análisis de co-palabras; redes bibliométricas; mapas científicos; diagramas estratégicos; visualización de la Información.

**Abstract:**
**Introduction**: Artificial Intelligence (AI) is having a significant impact in the field of communication, causing transcendental changes in the processing and consumption of information. The objective of this work was to analyze the most influential AI topic areas in the field of communication based on scientific literature. **Methodology**: 996 references indexed in Web of Science between 2004-2024 were selected, a bibliometric analysis of co-words was carried out and visualization techniques were applied to build scientific maps. **Results:** The most relevant thematic areas were datafication, the linking of AI with social media and digital journalism. The emerging area of generative AI was identified, linked to new AI models, such as ChatGPT, designed to generate content in the form of written text, audio, images or videos. Another emerging topic area was China's impact on the use of AI in communication. **Discussions:** Despite the impact of AI in communication, the field is still in the process of structuring, with few consolidated topics. **Conclusions**: This study made it possible to identify the thematic areas of the field studied, as well as the detection of emerging trends.

## 1. Introducción

La IA (Inteligencia Artificial) se define como la capacidad de las máquinas para aprender y demostrar inteligencia, que contrasta con la inteligencia humana (Teigens *et al.*, 2020). Los avances de la IA en cálculo y procesamiento de grandes cantidades de datos han permitido a los ordenadores realizar con éxito tareas de aprendizaje cada vez más complejas, lo que ha revolucionado la forma en que nos comunicamos. La IA también es una herramienta para poder gestionar y analizar el inmenso volumen de datos que se genera actualmente en las redes sociales y en Internet. La IA comprende una variedad de herramientas (tales como *Machine Learning*, *Deep Learning* y *Natural Language Processing*) por su capacidad para la extracción, análisis y uso de información para la toma de decisiones (Rangel, 2022).

En general, la IA está transformando muchos aspectos de la vida social, política y económica (Túñez, 2021), se habla incluso de una cuarta revolución industrial que transformará empresas y redefinirá paradigmas de investigación (Teigens *et al.*, 2020). La IA es una disciplina perteneciente al ámbito de las ciencias de la computación que propone métodos y técnicas para el desarrollo de programas informáticos con la habilidad de aprender y razonar como hacemos los humanos. Los robots de IA y los agentes conversacionales inteligentes han despertado un gran interés social debido a su impresionante capacidad para componer historias y ensayos, resolver problemas de programación y proporcionar respuestas concisas a preguntas.

La IA ha tenido un impacto trascendental el ámbito de la comunicación (Herrera-Ortiz *et al.*, 2024). La integración de la IA en el campo de la comunicación está cambiando la forma en la que se generan contenidos, han aparecido nuevas plataformas capaces de producir artículos, informes e incluso noticias completas de manera automatizada. La comunicación se ha transformado de un proceso lineal a una relación sistémica, de todos con todos, en el que el discurso global se va modificando a medida que se difunde por la interacción de los usuarios organizados en redes de distribución.

Las plataformas digitales fomentan que las personas sean agentes activos compartiendo información y siendo autores de sus propios contenidos (Shah *et al.*, 2020). Esa interacción ha provocado un aumento de canales, plataformas, terminales y tecnologías en Internet, así como la generación de un gran volumen de datos que sólo se pueden analizar con herramientas específicas de la IA (Perakakis *et al.*, 2019; Huang y Rust, 2020). Se espera que la IA siga evolucionando, ofreciendo capacidades aún más avanzadas, tanto en el procesamiento del lenguaje como en el análisis de datos o la toma de decisiones automáticas. Son muchas las investigaciones que reflejan el impacto que está teniendo la IA en el campo de la comunicación (Flew *et al.*, 2012; Karlsen y Stavelin, 2014; Hansen *et al.*, 2017; Lindén, 2017; Marconi y Siegman, 2017; Túñez-López *et al.*, 2018; Salazar, 2018; Vállez y Codina, 2018; Wölker y Powell, 2018; de Bustos e Izquierdo-Castillo, 2019; Segarra-Saavedra *et al.*, 2019; Túñez-Lopez *et al.*, 2019; Gran *et al.*, 2020; Ufarte *et al.*, 2020; Pérez y Perdomo, 2024).

Este artículo tiene como objetivo analizar la producción científica relacionada con el impacto de la IA en el campo de la comunicación para abordar las siguientes preguntas de investigación: 1) ¿cuáles son las principales áreas temáticas que utilizan la IA en el campo de la comunicación?; 2) ¿cómo se clasifican las áreas temáticas de la IA en el campo de la comunicación?; y 3) ¿cuáles son las tendencias de investigación de la IA en el campo de la comunicación?

## 2. Metodología

La metodología utilizada se basó en la aplicación de análisis de co-palabras y técnicas de visualización de información. El análisis de co-palabras se puede realizar a partir de un conjunto de documentos representativos de la producción de un área científica. Esta técnica parte del principio según el cual una especialidad de investigación puede ser identificada por su propio vocabulario. A partir del cómputo de las apariciones conjuntas de palabras, representadas como relaciones, se construyen redes de palabras que muestran la estructura conceptual y temática del campo de investigación analizado. En la primera fase, con las palabras seleccionadas, de una muestra obtenida de forma automática en una base de datos, se construyen matrices de co-ocurrencias de palabras. Una vez obtenida las matrices, se aplican diferentes tipos de análisis, de tal forma que la medida del enlace entre dos palabras será proporcional a la co-ocurrencia de esas dos palabras en el conjunto de documentos que se ha tomado como muestra (Callon *et al.*, 1986; 1991; Ding *et al.*, 2001). El procedimiento seguido se desarrolló en las siguientes etapas: recopilación de datos, selección de las unidades de análisis, creación de redes bibliométricas y construcción de mapas científicos o diagramas estratégicos. Para el procesamiento estadístico de los datos, la obtención de las redes bibliométricas y la construcción de los mapas científicos se utilizó la herramienta informática *RStudio Bibliometrix* (Aria y Cuccurullo, 2017). El análisis de las palabras asociadas, o análisis de co-palabras (Callon *et al.*, 1986; 1991) se dirige a la creación de redes bibliométricas y diagramas estratégicos donde se posicionan las áreas temáticas. Esta técnica parte del principio según el cual una especialidad de investigación puede ser identificada por su propio vocabulario. A partir del cómputo de las apariciones conjuntas de palabras, representadas como relaciones, se construyen redes de términos que muestran la estructura temática de los campos de investigación analizados, este procedimiento se ha utilizado en diferentes investigaciones (Callon *et al.*, 1983; Cobo *et al.*, 2011; 2014; Corrales-Garay *et al.*, 2024; Ding *et al.*, 2001; Leydesdorff y Welbers, 2011; López-Fraile *et al.*, 2023; Luo *et al.*, 2022; Montero-Díaz *et al.*, 2018; Peters y Van Raan, 1993; Raeeszadeh *et al.*, 2018; Segado-Boj *et al.*, 2023).

*2.1. Recopilación de datos y selección de unidades de análisis*

La fuente de información utilizada fue la base de datos *Web of Science* (WoS). La elección de WoS se debió a que proporciona numerosas herramientas de análisis para procesar los datos y ofrece información de producción científica altamente precisa. Dentro de la colección principal de WoS, los datos se obtuvieron de las bases de datos *Science Citation Index Expanded* (SCI-EXPANDED), *Social Sciences Citation Index* (SSCI), *Arts & Humanities Citation Index* (A&HCI) y *Emerging Sources Citation Index* (ESCI). La estrategia de búsqueda consistió en seleccionar el campo tema (Topic) = ("*artificial intelligence*" or IA), en el campo año de publicación (*Publication Year*) = "2004-2024", y en el campo categoría científica de WoS (*Web of Science Categories*) = "*Communication*".

En la base de datos WoS, los registros incluyen dos tipos de palabras-clave: palabras-clave de autor (*Author's keywords*), proporcionadas por los propios autores, y palabras-clave (*Keywords Plus*), que son términos del índice generados automáticamente a partir de los títulos de los artículos citados. Para el análisis bibliométrico, se seleccionaron las palabras-clave asignadas por los autores de los documentos por ser más específicas que las palabras-clave extraídas de forma automática. A continuación, se realizó una fase pre-procesamiento para normalizar las palabras-clave (unificando duplicados, palabras similares o escritas erróneamente y fusionando los plurales y singulares).

*2.2. Creación de redes bibliométricas de palabras-clave*

Las relaciones de co-ocurrencia de palabras-clave se representaron como redes de términos, en los que los nodos representaron las palabras-clave y las relaciones entre ellas se representaron como enlaces. La relación de co-ocurrencia se da entre dos palabras-clave que aparecen conjuntamente en un documento, es decir, si las palabras-clave *i* y *j* aparecen en el mismo documento, se considera que existe una relación de co-ocurrencia entre ellas. A su vez, esta relación puede cuantificarse, de forma que la relación represente, además, el número de documentos en los que las palabras-clave mencionadas aparecen conjuntamente. La fuerza de los enlaces depende del número de apariciones conjuntas en un mismo documento, esto es, si se dieran muchas co-ocurrencias sobre una misma palabra-clave, darían lugar a una alianza estratégica entre documentos, que se asoció con un tema de investigación.

Una vez creadas las redes bibliométricas de palabras-clave se calculó la intensidad de las asociaciones entre palabras, para ello fue necesario normalizar los enlaces. En este trabajo, se aplicó el índice de equivalencia (Callon *et al.*, 1991) como medida de similitud. La aplicación del índice de equivalencia estableció un peso a cada palabra-clave, equivalente a su importancia en el conjunto de documentos analizados. El cálculo de todos los coeficientes entre todos los pares de palabras-clave posibles generó un número excesivo de enlaces (demasiados para poder representarlos gráficamente), por eso el análisis se realizó con una versión simplificada que se configuró en función de la frecuencia mínima en el corpus, como mecanismo de poda de datos. Se seleccionaron solo las palabras-clave de autor que aparecieran con una frecuencia ≥ 3.

*2.3. Construcción del mapa científico o diagrama estratégico*

Para la construcción del mapa científico, o diagrama estratégico, se aplicaron técnicas de agrupamiento (*clustering*) (Börner *et al.*, 2003). Las técnicas de agrupamiento dividieron el conjunto de las redes bibliométricas de palabras-clave en diversos subconjuntos, los cuales debieron cumplir la condición de tener una gran cohesión interna. Los algoritmos de agrupamiento aplicados a las redes bibliométricas identificaron las subredes que formaron la red bibliométrica global, esto es, aquellos conjuntos de nodos que estuvieron fuertemente enlazados entre sí, pero débilmente enlazados con el resto de la red. Cada subred se asimiló a un área temática de investigación (cuya etiqueta estuvo identificada por las palabras-clave con mayor peso, dentro de cada área temática). Entre los métodos de agrupamiento de palabras-clave se utilizó el algoritmo del centro simple (Coulter *et al.*, 1998), que permitió crear subredes de palabras-clave vinculadas entre sí, que correspondieron a las áreas temáticas o los focos de interés. La aplicación de algoritmos de agrupamiento a la red bibliométrica permitió la creación de subredes con nodos fuertemente enlazados entre sí.

A continuación, para la clasificación de subredes se aplicaron dos métricas (Callon *et al.*, 1991; Coulter *et al.*, 1998):

1. *Centralidad de Callon*. Mide el grado de relevancia, a través de la interacción (o fuerza de los enlaces externos) de una red con respecto a otras redes. Se define como $c = 10 * \sum e_{kh}$, siendo $k$ una palabra-clave perteneciente a ese mismo tema, y siendo $h$ una palabras-clave perteneciente a otro tema.
2. *Densidad de Callon*. Mide el grado de desarrollo, a través de la intensidad interna de la red, o densidad de todos los enlaces entre las redes de palabras-clave. Se define como $d = 100 (\sum e_{ij} / w)$, donde $i$ y $j$ son palabras-clave pertenecientes al tema y $w$ el número de palabras-clave, nodos, que forman el tema.

Las subredes temáticas obtenidas se visualizaron en un mapa científico o diagrama estratégico. Los diagramas estratégicos constituyen visualizaciones bidimensionales de las redes bibliométricas ubicadas en cuatro cuadrantes, de acuerdo con sus valores de centralidad (grado de relevancia) y de densidad (grado de desarrollo) (Callon *et al.*, 1991). La posición estratégica de las redes temáticas, en los respectivos cuadrantes, proporcionó una lectura comprensible e interpretable del grado de importancia y del nivel de desarrollo de los temas (o áreas temáticas) que la componen. En el diagrama estratégico los temas se clasificaron en cuatro grupos (Figura 1):

- Cuadrante 1. Temas Motores (centrales y desarrollados). En este cuadrante se situaron los temas que presentaron una centralidad fuerte y una alta densidad. Configuraron los temas importantes, que constituyeron el centro del campo.
- Cuadrante 2. Temas Básicos (centrales y no desarrollados). En este cuadrante se situaron los temas centrales e importantes para el desarrollo del campo (susceptibles de convertirse en centrales y desarrollados, y, por tanto, desplazarse al cuadrante 1).
- Cuadrante 3. Temas Nicho (periféricos y desarrollados). En este cuadrante se situaron los temas con fuerte intensidad de sus relaciones internas (gran densidad), vinculados con temas de investigación bien desarrollados, especializados pero cuyos enlaces externos fueron débiles.
- Cuadrante 4. Temas Emergentes (periféricos y no desarrollados). En este cuadrante se ubicaron los temas con una centralidad y densidad baja, correspondieron tanto a los temas emergentes (o temas periféricos susceptibles de desarrollarse en el futuro) como a los temas marginales (o temas periféricos susceptibles de desaparición).

**Figura 1.**

*Distribución de los temas en los cuadrantes del diagrama estratégico*

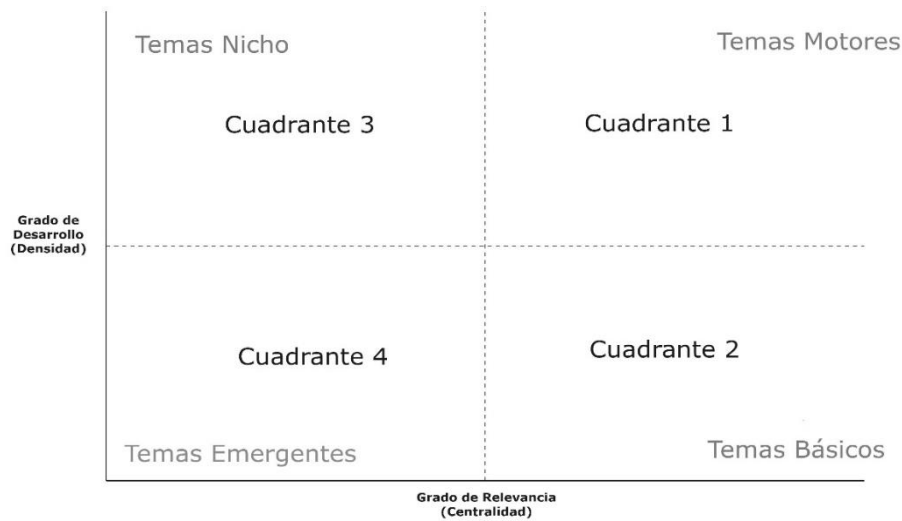

**Fuente:** Elaboración propia (2024).

Las visualizaciones se enriquecieron añadiendo una tercera dimensión a los elementos representados, como fue el número de documentos asociados al tema. Se configuró un número mínimo de palabras-clave por clúster (tres) para que se establecieran visualizaciones efectivas de los datos. Las áreas temáticas se mapearon como esferas de redes temáticas, en las que el volumen fue proporcional al número de documentos vinculados a ese tema. Cada red temática, o tema detectado, se etiquetó usando el nombre de la palabra-clave más significativa de la red.

Según la distribución de las redes temáticas en los cuadrantes, las áreas temáticas se pudieron clasificar en tres categorías (Callon *et al.*, 1991) (Figura 2):

- Categoría 1. La distribución de los temas en la primera bisectriz (cuadrante 1 - cuadrante 4) indica que el campo se organiza en torno a un núcleo de temas bien desarrollados, con los cuales se relaciona una serie de temas periféricos y poco desarrollados.
- Categoría 2. La distribución de los temas alrededor de la segunda bisectriz (cuadrante 2 - cuadrante 3) indica un campo en vías de estructuración, pocos temas están en la posición central y la red es muy policéntrica.
- Categoría 3. La distribución de los temas en los diferentes cuadrantes caracteriza un campo de estructura importante, consolidada, compleja y rica, con temas motores, básicos, en desarrollo, emergentes o en declive.

**Figura 2.**

*Categorías de los diagramas estratégicos*

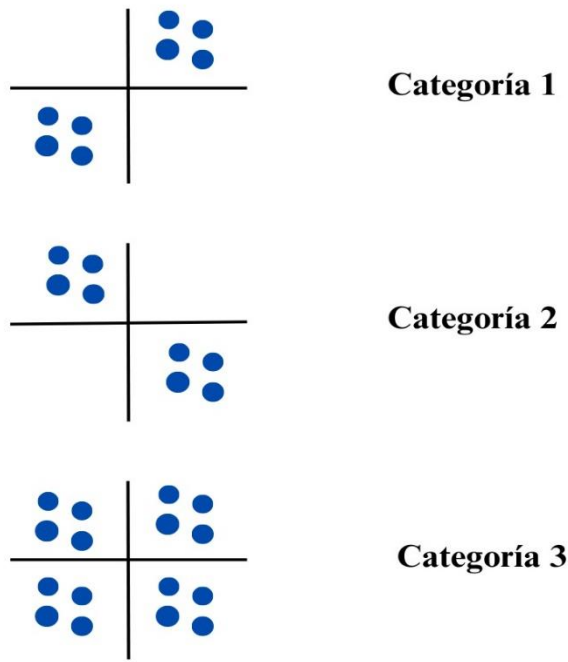

**Fuente:** Elaboración propia (2024).

Además, el seguimiento de las posiciones de los temas en el diagrama estratégico sigue un movimiento circular, que permite predecir el posible comportamiento de las redes en el futuro (Courtial, 1990): las áreas temáticas situadas en el cuadrante 2 presentan una tendencia a moverse al cuadrante 1 (esto significa que los temas centrales en el mapa tienen tendencia a desarrollarse internamente) y las áreas temáticas situadas en el cuadrante 1 tienden a alejarse de las líneas de investigación dirigiéndose hacia el cuadrante 3. Los movimientos circulares de los temas en el diagrama estratégico se simbolizan de la forma siguiente: los temas nacen en el cuadrante 4 (temas emergentes o en declive), pasan al cuadrante 2 (temas básicos), continúan dirigiéndose al cuadrante 1 (temas motores) y terminan su avance en el cuadrante 3 (temas nicho).

## 3. Resultados

Se recuperaron un total de 996 documentos en la base de datos WoS. Sus características específicas se presentaron en la Tabla 1.

**Tabla 1.**

*Información general de la muestra analizada*

| Periodo de tiempo analizado | 2004-2024 |
|---|---|
| **Información principal** | |
| Fuentes (Revistas, Libros, otros) | 251 |
| Documentos | 996 |
| Tasa de crecimiento anual | 28,07 |
| **Contenido del documento** | |
| Palabras-clave Plus (*Keywords Plus*) | 940 |
| Palabras-clave del Autor (*Author's Keywords*) | 2.882 |

**Fuente:** Elaboración propia (2024).

Se obtuvieron un total de 2.882 palabras-clave de autor (*Author's Keywords*). Se seleccionaron sólo aquellas palabras-clave cuya frecuencia fuera ≥ 3. El resultado fue una muestra de 387 palabras-clave (si se hubieran seleccionado las palabras-clave con una frecuencia menor, la red resultante hubiera sido difícil de interpretar por la gran cantidad de nodos y enlaces obtenidos). Las veinte palabras-clave más frecuentes se muestran en la Tabla 2.

**Tabla 2.**

*Las 20 palabras-clave de autor más frecuentes*

| Palabras-clave de autor | Frecuencias |
|---|---|
| *artificial intelligence* | 501 |
| *Ai* | 61 |
| *Journalism* | 59 |
| *Algorithms* | 53 |
| *social media* | 49 |
| *machine learning* | 48 |
| *Automation* | 35 |
| *Communication* | 35 |
| *Ethics* | 35 |
| *big data* | 34 |
| *Chatgpt* | 28 |
| *disinformation* | 27 |
| *fake news* | 27 |
| *automated journalism* | 26 |
| *human-machine communication* | 24 |
| *technology* | 23 |
| *chatbots* | 18 |
| *innovation* | 18 |
| *fact-checking* | 17 |
| *natural language processing* | 14 |

**Fuente:** Elaboración propia (2024).

Se contabilizaron todas las apariciones conjuntas y las relaciones se representaron en una red bibliométrica de palabras-clave, en la que la existencia de las relaciones y su fuerza de asociación dependió del número de co-ocurrencias entre los términos en los diferentes documentos considerados (Figura 3).

**Figura 3.**

*Red bibliométrica de palabras-clave*

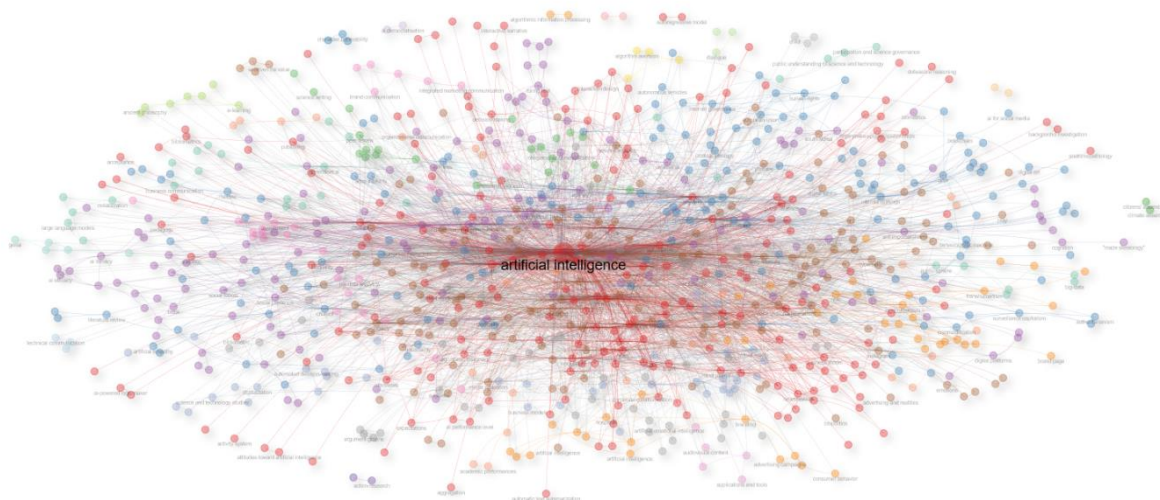

**Fuente:** Elaboración propia (2024).

A partir de la red bibliométrica de palabras-clave, se aplicaron algoritmos de agrupamiento y se creó el diagrama estratégico mediante técnicas de reducción de la dimensionalidad, y utilizando las medidas de centralidad y densidad. Se identificaron 12 grandes áreas temáticas (Tabla 3).

**Tabla 3.**

*Redes temáticas detectadas en el diagrama estratégico*

| Etiquetas de las redes temáticas | Número de documentos | Centralidad de Callon | Densidad de Callon |
|---|---|---|---|
| ARTIFICIAL INTELLIGENCE | 1.264 | 3,6 | 19,25 |
| SOCIAL MEDIA | 334 | 1,5 | 24 |
| DATAFICATION | 181 | 0,9 | 25 |
| AI (ARTIFICIAL INTELLICENCE) | 174 | 0,7 | 23 |
| INTERNET | 78 | 0,75 | 27,7 |
| LABOR | 49 | 0.35 | 39 |
| CHINA | 32 | 0,22 | 26,8 |
| DIGITAL JOURNALISM | 29 | 0,24 | 21,5 |
| GENERATIVE AI | 20 | 0,11 | 18,6 |
| DATA SCIENCE | 9 | 0,26 | 25 |
| ASSESMENT | 8 | 0 | 25 |
| E-LEARNING | 6 | 0,075 | 16,6 |

**Fuente:** Elaboración propia (2024).

En el diagrama estratégico, las redes temáticas obtenidas se posicionaron en los cuadrantes en función de los indicadores de centralidad (nivel de relevancia) y de densidad (nivel de desarrollo) (Figura 4), distinguiéndose: temas motores, temas básicos, temas nicho y temas emergentes o marginales.

**Figura 4.**

*Posicionamiento de los temas de investigación en el diagrama estratégico*

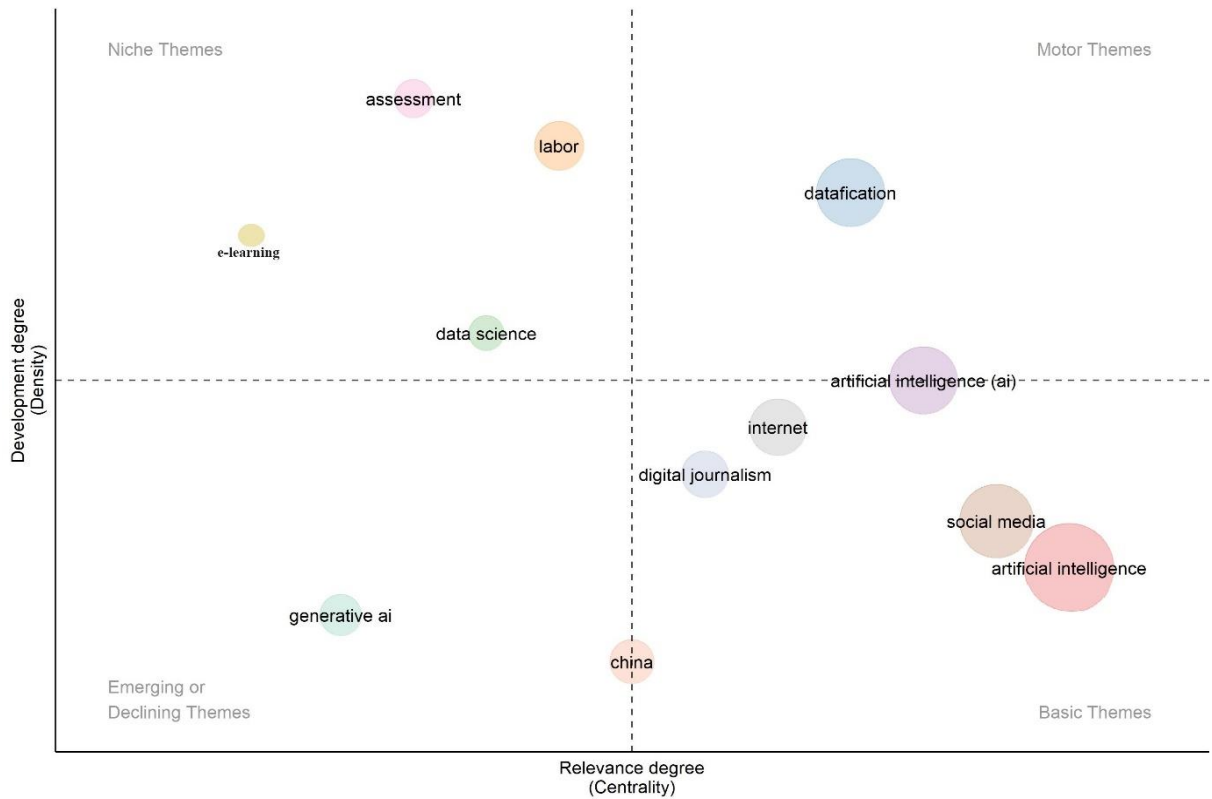

**Fuente:** Elaboración propia (2024).

## 3.1. Temas motores (Cuadrante 1)

En este cuadrante se situaron las áreas temáticas centrales y desarrolladas (con una elevada centralidad y densidad). Se detectó un tema motor etiquetado con la palabra-clave: "Datafication", que integró 181 documentos. La datificación consiste en tratar la información hasta convertirla en datos, este proceso permite medir, almacenar y consultar posteriormente estos datos mediante herramientas tecnológicas de IA. La datificación implica la conversión de información en datos digitales que pueden ser objeto de cuantificación, procesamiento y análisis mediante algoritmos y herramientas computacionales. En relación con la datificación, el periodismo de datos (*big data*) constituye el nuevo contexto de la IA en el campo de la comunicación, se trata de una forma de periodismo en la que los periodistas utilizan grandes conjuntos de datos para investigar y contar historias (como el análisis de datos y la extracción de información relevante).

La datificación está vinculada también a la comunicación estratégica y a la segmentación de audiencias. La capacidad de la IA para analizar grandes cantidades de datos y predecir patrones de comportamiento se vincula a la posible manipulación de la información en el campo de la comunicación (como los algoritmos de recomendación que pueden influir en las opiniones de las personas). En la comunicación estratégica destaca la aplicación de iniciativas de comunicación para analizar, conectar y construir relaciones con las audiencias, así como la vinculación con procesos de comunicación corporativa, responsabilidad social, medios digitales y sociales para lograr metas organizacionales. El área temática de la datificación integró otras subredes temáticas relacionadas con los *chatbots*, los medios digitales y la seguridad en las comunicaciones (los *chatbots*, basados en IA, son programas que pueden interactuar con una persona real siguiendo determinados patrones, capaces de mantener una conversación en tiempo real por texto o por voz, como los *chatbots* de atención al cliente, y los asistentes virtuales, como *Siri* o *Alexa*).

*3.2. Temas básicos (Cuadrante 2)*

En este cuadrante se posicionaron los temas centrales, pero no desarrollados. Giró en torno a cinco grandes áreas temáticas: "Artificial Intelligence", "Social Media", "(AI) Artificial Intelligence", "Internet" y "Digital Journalism". El tema básico "Artificial Intelligence" agrupó el mayor número de documentos publicados, 1.264. El tema "Social Media" congregó 334 documentos. El siguiente tema básico fue "(AI) artificial intelligence", que integró 174 documentos. Los otros temas básicos identificados fueron "Internet" con 78 documentos y "Digital Journalism" con 29 documentos. En este cuadrante se situaron los temas básicos para comprender las capacidades, aplicaciones y desafíos asociados con la IA y su interacción con la comunicación. La IA facilita la creación de noticias falsas y virales (*fake news*), esto constituye uno de los problemas a los que se enfrentan los profesionales de la comunicación como es la verificación de la autenticidad de la información generada automáticamente.

Otro de los temas básicos estuvo vinculado a las redes sociales, referido a todas las redes y medios que han surgido en los últimos años con el Internet, provocando una transformación digital en la prensa tradicional, que ha migrado en gran medida a plataformas digitales, permitiendo una mayor interactividad con los lectores y la posibilidad de actualizar las noticias en tiempo real. También entre los temas fundamentales surgió la ética, que se ocupa de cómo la IA debe comportarse y cómo debe ser utilizada. En este contexto, se espera que los profesionales de la comunicación no solo informen, sino que también comprendan y utilicen herramientas de IA para mejorar la calidad de su trabajo. Los sistemas inteligentes presentan oportunidades para la innovación, pero también desafíos en la preservación de la ética.

*3.3. Temas nicho (Cuadrante 3)*

En este cuadrante se encontraron los temas periféricos y desarrollados del campo. La fuerte intensidad de sus relaciones internas (gran densidad) hizo que se correspondieran con problemáticas de investigación cuyo estudio estuvo bien desarrollado, pero cuyos enlaces externos fueron débiles, es decir, interactuaron débilmente con respecto a otras subredes. Este cuadrante giró en torno a cuatro temas: "*Labor*", "*Data science*", "*Assesment*" y "*E-Learning*". El tema nicho "*Labor*" integró 49 documentos. El tema "*Data science*" agrupó 9 documentos. El tema "*Assesment*" estuvo formado por 8 y el tema "*E-Learning*" estuvo constituido por 6 documentos. En este cuadrante se situaron los temas especializados y los temas relacionados con las aplicaciones de la IA en la comunicación, tales como la obtención de información a partir de grandes cantidades de datos, como un recurso en la toma de decisiones, o el aprendizaje electrónico, para mejorar la experiencia de aprendizaje y comunicación.

*3.4. Temas emergentes o declive (Cuadrante 4)*

En este cuadrante se ubicaron los temas periféricos y no desarrollados. Se detectaron dos temas: "*Generative AI*" y "China". El área temática emergente etiquetada con la palabra-clave "*Generative AI*" integró 20 documentos y el área temática "China" agrupó a 32 documentos. La IA generativa, que constituyó un importante tema emergente en IA, se refiere a los nuevos modelos de IA diseñados para generar nuevo contenido en forma de texto escrito, audio, imágenes o videos. La IA generativa está transformando las formas de producción de los contenidos digitales. La aparición de herramientas de creación de texto, como ChatGPT, y la creación de imágenes y vídeos a partir de texto han supuesto un cambio paradigmático en la forma en que los profesionales de la comunicación afrontan el proceso creativo y el acceso al conocimiento. ChatGPT surgió como un *chatbot* impulsado por IA que proporciona respuestas inteligentes a las consultas de los usuarios (Halaweh, 2023). Esta innovación ha provocado que la investigación, vinculada a la tecnología y la comunicación, se plantee nuevos retos sobre el uso de la IA generativa. Las aplicaciones y los casos de uso de la IA generativa son muy amplios (como crear una historia corta basada en el estilo de un autor en particular, componer una sinfonía en el estilo de un compositor famoso o crear un videoclip a partir de una descripción textual simple).

En cuanto al tema emergente "China", el diagrama reflejó la importante repercusión de este país en el uso de la IA como instrumento de comunicación de masas. China posee un número enorme de patentes de IA y sus académicos han publicado un número muy elevado de investigaciones en IA. China está planteando un cambio de paradigma de los medios de comunicación, en el que los periodistas cuentan con medios que utilizan IA y algoritmos para mejorar la calidad de sus contenidos.

## 4. Discusión

En este trabajo se ha identificado la estructura temática de la IA en el campo de investigación de la comunicación. La metodología utilizada fue el análisis de palabras asociadas, una técnica aplicada al análisis de contenido que se dirige a la representación de las estructuras y relaciones de un dominio de conocimiento, por medio del análisis del número de veces que co-ocurren simultáneamente dos términos en los documentos. Para ello se descargaron los datos desde la base de datos WoS. Los resultados permitieron la representación de las estructuras y patrones semánticos que subyacen ocultos en los grandes volúmenes de datos. Las redes temáticas detectadas nos permitieron analizar la estructura y clasificar los temas, a partir de la posición de las redes temáticas en el diagrama estratégico. Cada red temática, o grupo de palabras-clave, se presentó como un círculo en el mapa científico. La etiqueta, o el nombre del círculo, estuvo determinada por la palabra-clave cuya asociación interna fue la más alta entre las sub-áreas temáticas en el grupo. Para posicionar las redes temáticas en el diagrama estratégico se aplicaron dos indicadores: centralidad (grado de relevancia) y densidad (grado de desarrollo). La centralidad midió las principales redes temáticas, o líneas de investigación, y la relevancia de los temas en el desarrollo global del campo de investigación analizado. La densidad proporcionó una idea del nivel de desarrollo de las redes temáticas. En función de la posición de las redes temáticas en los diferentes cuadrantes del diagrama estratégico se interpretaron los valores obtenidos.

En total se identificaron doce grandes redes temáticas, correspondientes a las doce grandes áras temáticas. Las redes temáticas ubicadas en los cuadrantes superior e inferior derechos fueron las de mayor relevancia e interés, ya que en estas posiciones se distinguieron seis redes temáticas importantes vinculadas al procesamiento de los datos, las nuevas plataformas digitales surgidas con Internet, los medios sociales y el periodismo digital: "*Datafication*", "*Artificial Intelligence*", "*Social Media*", "*(AI) Artificial Intelligence*", "Internet" y "*Digital Journalism*".

Las redes temáticas ubicadas en los cuadrantes superior e inferior izquierdos se identificaron con redes temáticas vinculas, por un lado, con temas especializados y aplicaciones específicas de la IA al campo de la comunicación ("*Labor*", "*Data science*", "*Assesment*", "*E-Learning*") y, por el otro, con temas emergentes como la IA generativa y China ("*Generative AI*" y "China"). Dentro de la IA generativa destacó la herramienta ChatGPT, un programa con capacidad para generar una gran variedad de contenido, desde artículos para un blog, hasta mensajes para redes sociales o respuestas a preguntas frecuentes, lo que la hace versátil y adaptable a diversas necesidades en el ámbito de la comunicación. En la actualidad, la IA generativa se utiliza para la creación de contenidos, la verificación de hechos, el procesamiento de datos, la generación de imágenes, la conversión de voz y la traducción, lo que reduce la carga de trabajo humano y aumenta su eficacia (Nishal y Diakopoulos, 2024). A pesar del enorme potencial de la IA generativa, la generación de textos, imágenes y audios artificiales difíciles de distinguir de contenidos reales puede ser usado para crear noticias falsas (*fake news*) y desinformar (*disinformation*) (Aïmeur *et al.*, 2023). El papel de la IA generativa en el periodismo también diluye el poder de los profesionales de los medios, cambia la producción de noticias tradicional y plantea múltiples cuestiones éticas (Shi y Sun, 2024).

Por último, la metodología utilizada nos permitió clasificar la estructura y morfología de la investigación en IA en comunicación. Las redes temáticas se posicionaron principalmente en los cuadrantes 2 y 3. Esta distribución se clasificó en la categoría 2 (segunda bisectriz). Una configuración de este tipo indicó un campo en vías de estructuración (esto es, un grupo de temas básicos y otro grupo de temas especializados) con mucho recorrido en el futuro. Además, los movimientos circulares de los temas, en el diagrama estratégico, mostró que el patrón de evolución dinámica lo constituyó la IA generativa, considerándose el nuevo desafío de la IA en campo de la comunicación.

## 5. Conclusiones

Este estudio permitió responder a las cuestiones de investigación formuladas al principio del trabajo. Con respecto a la primera pregunta de investigación, se identificaron los principales ejes temáticos: "*Datafication*", "*Artificial Intelligence*", "*Social Media*", "Internet", "*Digital Journalism*". Con respecto a la segunda pregunta de investigación, la estructura semántica detectada se clasificó en la categoría 2, indicando que la organización del campo de conocimiento se encuentra en vías de estructuración. Con respecto a la tercera pregunta, la tendencia de investigación se dirige a la IA generativa. Según este resultado se pudo deducir que el tema de la IA generativa constituye el siguiente reto en la IA, revelando que se trata de un tema que experimentará, con mucha probabilidad, un desarrollo exponencial en el futuro inmediato. Como tema emergente también destacó la repercusión de China en el uso de la IA como instrumento de comunicación de masas. En síntesis, el mapa científico resultó un instrumento eficaz para proporcionar una representación de la investigación de la IA en la comunicación. Aunque este estudio no estuvo exento de algunas limitaciones. En el análisis realizado solo se consideró la producción científica incluida en la colección de WoS, lo que podría excluir potencialmente un número reducido de artículos. En próximos trabajos, se podría ampliar las bases de datos utilizadas.

## 6. Referencias

**AUTORA:**

**Carmen Gálvez:**
Universidad de Granada.

Profesora Titular del Departamento de Información y Comunicación de la Universidad de Granada, desde 2009. Pertenece al grupo de investigación HUM466 −Acceso y evaluación de la información científica−. Áreas de interés: Análisis de co-palabras, Análisis de Redes Sociales (ARS), Normalización de términos en bases de datos documentales. En la actualidad su investigación se centra en: análisis de las estructuras conceptuales y análisis de co-palabras. Profesora en el Máster Información y Comunicación Científica de la Universidad de Granada, asignatura: Visualización de la Información: Modelo Metodológico del Análisis Estructural y de Redes Sociales.
cgalvez@ugr.es

**Índice H:** 11
**Orcid ID:** https://orcid.org/0000-0001-7454-1254
**Scopus ID:** https://www.scopus.com/authid/detail.uri?authorId=36723249300
**Web of Science:** https://www.webofscience.com/wos/author/record/E-8212-2011